\pdfoutput=1
\documentclass[journal=nalefd,manuscript=letter,layout=twocolumn]{achemso}

\usepackage[version=3]{mhchem} 
\usepackage{hyperref}
\usepackage{epstopdf}

\author{S.M. Hornett}
\email{s.m.hornett@exeter.ex.ac.uk}
\author{R. I. Stantchev}
\author{M.Z. Vardaki}
\author{C. Beckerleg}
\author{E. Hendry}
\affiliation[Exeter University]
{Physics Building, University of Exeter, Exeter, Devon, UK}

\title[]
  {Subwavelength THz imaging of graphene photoconductivity}

\abbreviations{}

\begin{document}
\begin{abstract}
Using a spatially structured, optical pump pulse with a THz probe pulse, we are able to determine spatial variations of the ultrafast THz photoconductivity with sub-wavelength resolution (75 $\mu m \approx \lambda/5$ at 0.8 THz) in a planar graphene sample. We compare our results to Raman spectroscopy and correlate the existence of the spatial inhomogeneities between the two measurements. We find a strong correlation with inhomogeneity in electron density. This demonstrates the importance of eliminating inhomogeneities in doping density during CVD growth and fabrication for photoconductive devices.
\end{abstract}

\section{Introduction}

The unique opto-electronic properties of graphene have received a lot of attention \cite{CastroNeto2009TheGraphene}. For example, ultrafast carrier relaxation\cite{George2008UltrafastGraphene,Dawlaty2008MeasurementGraphene,Choi2009BroadbandGraphene,Wang2010UltrafastGraphene,Hale2011HotGraphene}, highly tunable doping levels\cite{Geim2007TheGraphene.}, theoretical mobilities in excess of 150,000 cm$^{2}$V$^{-1}$s$^{-1}$ \cite{Geim2007TheGraphene.}, and high thermal conductivity \cite{Balandin2008SuperiorGraphene} all lend themselves to a number of interesting device applications.  However, the large scale manufacturing of this 2D material, usually through chemical vapour deposition (CVD), is not yet perfected and it is well established that there are a number of sources of quality degrading, spatial inhomogeneities such as charge puddles, grain boundaries, substrate induced strain variations, surface impurities, multilayer nucleation sites and fabrication residues
\cite{Zhang2009OriginGraphene,Gammelgaard2014GrapheneTreatments,Yu2011ControlDeposition,EunLee2012OpticalGraphene}. Raman spectroscopy and imaging has emerged as an important characterisation tool, due to its sensitivity to layer number \cite{Ferrari2006RamanLayers}, strain\cite{EunLee2012OpticalGraphene}, carrier concentration\cite{Das2008MonitoringTransistor} and defects\cite{M.S.DRESSELHAUS2010DefectSpectroscopy.,Canccado2011QuantifyingEnergies,Eckmann2012ProbingSpectroscopy,Froehlicher2015RamanScattering}. However, the photoconductivity, a critical quantity for many opto-electronic applications including photodetectors \cite{Liu2014GrapheneTemperature,Koppens2014PhotodetectorsSystems}, cannot be explicitly determined directly in Raman due to the limited number of observable quantities.

Given its relevance to many opto-electronic applications of graphene, optical-pump THz-probe spectroscopy has attracted considerable interest in the literature in recent years \cite{Docherty2012ExtremeGases.,Tielrooij2013PhotoexcitationGraphene,Freitag2013PhotoconductivityGraphene,Frenzel2014Semiconducting-to-metallicGraphene,Jensen2014CompetingGraphene}. This experiment can determine the ultrafast photoconductivity of graphene, which is now understood to have a complex dependence on mobility, electron concentration and relaxation rate\cite{Frenzel2014Semiconducting-to-metallicGraphene,Jensen2014CompetingGraphene,Shi2014ControllingGating}. However, due to the large THz spot sizes ($\approx$ mm) used in these experiments, they typically provide spatially averaged information, and are therefore ignorant of the small spatial inhomogeneities typical in CVD graphene. 

In this study, we introduce a technique able to directly image how these spatial inhomogeneities affect the local, photoconductive THz response of graphene. This is achieved via spatial patterning of the optical pump beam, allowing us to selectively sample our graphene, and thereby building a THz photoconductivity map of our CVD sample. We compare the spatially dependent THz photoconductivity to Raman spectral maps and find there to be various correlated features. We find that small regions of graphene with low electron density display a strongly suppressed photoconductivity on ultrafast timescales. Since the resolution of our measurement is determined by the patterned optical pump pulse, we are able to observe these small regions of suppressed THz photoconductivity on markedly sub-wavelength length scales ($75$ $\mu$m $\approx \lambda/5$ at 0.8 THz).

\begin{figure}
	\includegraphics{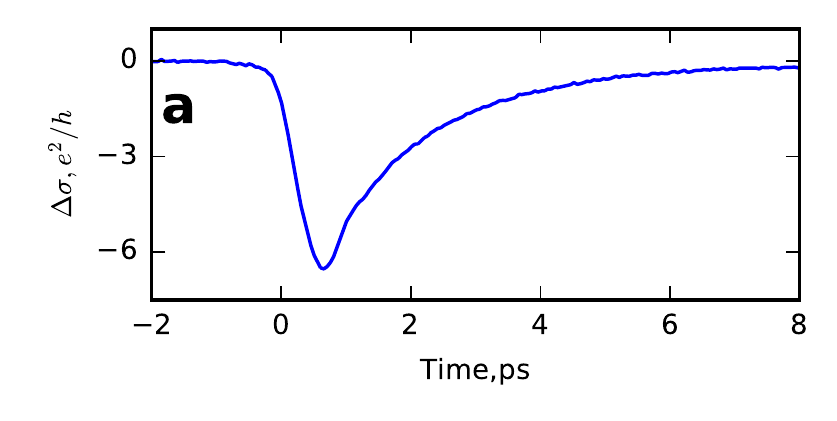}
    \includegraphics[]{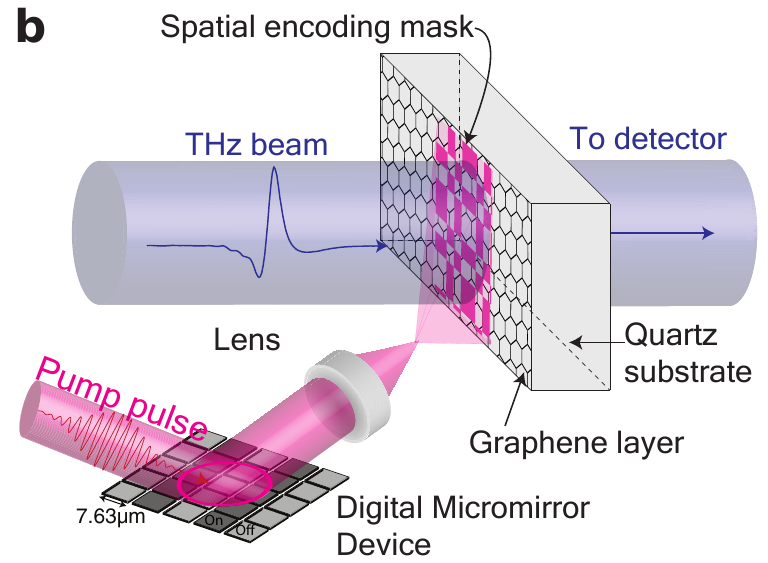}
	\caption{(a) Spatially averaged photoconductivity as a function of time delay after photo-excitation at 0 ps. (b) The imaging setup; a patterned 800 nm pump beam is used to photoexcite a graphene sample on quartz substrate (from \href{https://graphene-supermarket.com/}{graphene supermarket}). The graphene is then probed with a THz pulse $(\lambda_0=$400 $\mu m)$. Note that the DMD used (DLP lightcrafter, Texas Instruments) has a 13$^{\circ}$ angle between the individual mirrors and the plane of the mirror array, which introduces a wavefront distortion to the excitation beam. In order to remove the temporal smearing arising from this, we photoexcite at an incident angle of 13$^{\circ}$ to normal. Greater detail of this experimental arrangement is shown in the supplementary materials.}
	\label{fig_diagram}
\end{figure}

We use an amplified femtosecond laser system (800 nm, 1 KHz repetition rate, $\sim $100 fs)  to generate and detect our THz probe beam in a pair of ZnTe crystals through optical rectification \cite{Ulbricht2011CarrierSpectroscopy} and balanced electro-optic sampling \cite{Ulbricht2011CarrierSpectroscopy}, respectively. This allows us to determine the electric field, \textit{E}, of a single cycle THz pulse (central frequency $\sim $0.8 THz,  $\text{FWHM}$ $\sim $1.0 THz\cite{Ulbricht2011CarrierSpectroscopy}) transmitted through our sample. Note that, in all the data presented here, we discuss only changes in the peak transmitted field, as in ref. \cite{Jensen2014CompetingGraphene} . This gives a spectrally averaged measurement weighted to the spectral wavelength of our THz pulse ($\lambda_0=$400 $\mu m$). The femtosecond laser system also provides a third pump beam used to photoexcite the graphene. Our raw measurement of the temporal photoexcitation dynamics of graphene are shown in figure \ref{fig_diagram}(a) where we plot  $\Delta E$, defined as 
\begin{equation}
\Delta E = E_{\text{Pump On}} - E_{\text{Pump Off}},
\label{eqn1}
\end{equation}
as we vary the time between the optical pump and THz probe pulses. Here, the photoexcitation pulse arrives at $\sim 0$ ps. We see a fast, sub-picosecond carrier rise time followed by picosecond relaxation times (associated with carrier cooling) as observed previously in refs.  \cite{Docherty2012ExtremeGases.,Tielrooij2013PhotoexcitationGraphene,Frenzel2014Semiconducting-to-metallicGraphene,Jensen2014CompetingGraphene}. From this measurement one can extract the photoconductivity, $\Delta \sigma$, via the relation \cite{Choi2009BroadbandGraphene}
\begin{equation}
\Delta \sigma = -\frac{1+n_{\text{sub}}}{Z_0} \frac{\Delta E}{E_{\text{Pump Off}}}, 
\label{eqn2}
\end{equation}
where $E_{\text{Pump Off}}$ is the transmitted THz field before photoexcitation, $Z_0$ the impedance of free space and $n_{\text{sub}}\approx 1.9$ is the THz refractive index of the quartz substrate. From the data in fig. 1(a), it is clear that we have a negative photoconductivity (i.e. a conductivity which decreases on photoexcitation). This is typical for graphene with an intrinsic Fermi level greater than 120 meV \cite{Shi2014ControllingGating.} (from Raman measurements \cite{EunLee2012OpticalGraphene}, we estimate the intrinsic Fermi level of our sample to be $\sim 550 $ meV). However, it is important to note that this is a spatially averaged result: due to the restrictive diffraction limit for THz radiation, THz photoconductivity can typically only be determined with $\sim $ mm spatial resolution. 

To overcome this resolution limit, we introduce spatial modulation in the optical pump beam, as illustrated in figure \ref{fig_diagram}(b). For this we employ a digital multi-mirror device (DMD) to pattern the incoming optical pump beam.The simplest spatial dependence which can be used is a single raster scanning spot. This is analogous to near field probes \cite{vonRibbeck2008SpectroscopicMicroscope}
or scanning apertures \cite{Macfaden20143mumMicroscopy}. However single apertures and scatterers produce tiny signals due to their small size with respect to the THz wavelength \cite{Stantchev2016NoninvasiveDetector}. To achieve optimum signal to noise, we therefore pattern our photoexcitation beam into binary intensity masks derived from Hadamard matrices \cite{Harwit1979HadamardOptics,Stantchev2016NoninvasiveDetector}, as explicitly described in Ref. \cite{Stantchev2016NoninvasiveDetector} and in the supplementary information. Knowledge of the masking patterns and the corresponding far-field detector readout is combined to obtain an image of the THz photoconductivity of the object, our CVD graphene sample. In this experimental design, the theoretical imaging resolution is limited by the Rayleigh criterion for our pump beam. However, in practice the signal to noise ratio in experiment leads to long measurement times for high resolution \cite{Stantchev2016NoninvasiveDetector} . We find that a resolution of 75 $\mu$m is sufficient to resolve most of the conductivity features in our sample.

\section{Results}
\begin{figure}
\includegraphics[width=\linewidth]{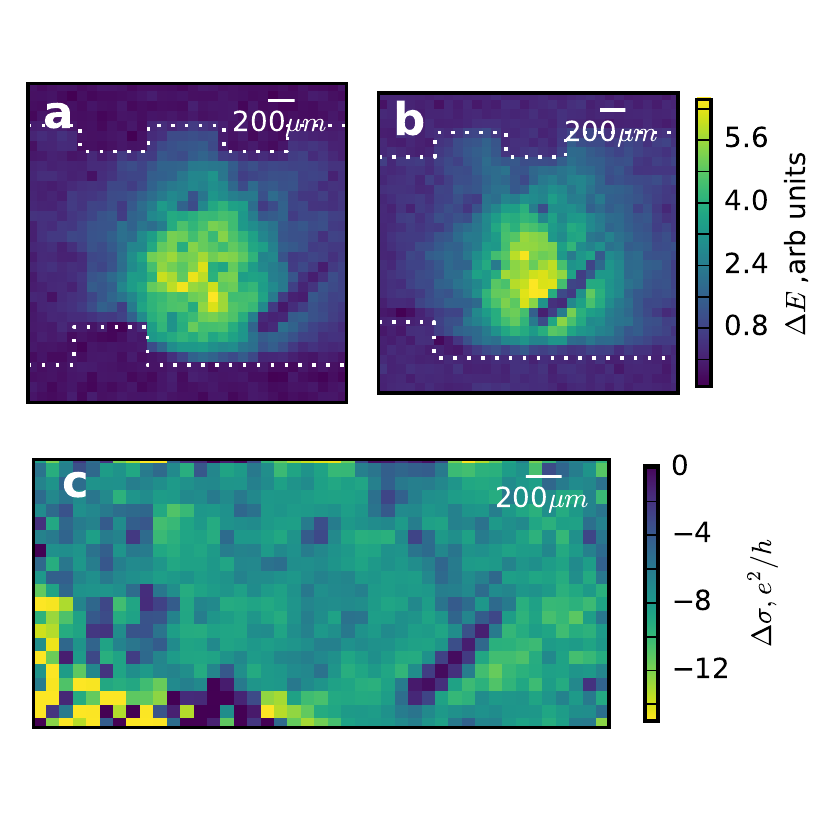}
\caption{(a) $\Delta E$ measured at $x=0\mu m$ showing the graphene response convoluted with the THz probe spot (white dotted lines shows gold alignment markers). (b)  $\Delta E$ measured with a shift of 450 $\mu$m with respect to (a). (c) Spatial dependence of the THz photoconductivity, as calculated following the procedure in text.}
\label{THz-Extraction}
\end{figure}

The imaging results are shown in figure \ref{THz-Extraction}. We measure at the peak in $\Delta \sigma$ shown in \ref{fig_diagram}a. Figures \ref{THz-Extraction}(a) and (b) are images showing the spatial dependence of $\Delta E$ as recorded with our single pixel imaging scheme. The THz probe beam profile can be observed in the centre of both images, with a number of additional features inside the spot. In order to separate the spatial response of the sample from any spatial inhomogeneities of the THz and optical pump beams, an averaging technique is employed, where the sample is laterally displaced with respect to the pump and probe beams. This allows us to extract any intensity variations associated with the graphene sample itself. In figure \ref{THz-Extraction}(b) the sample has been horizontally offset by 450 $\mu m$. As the sample translates left to right, we use gold markers (square features at top and bottom) to track its movement. The full set of results are shown in supplementary materials as video S1. The average beam profile is then extracted by taking the mean of all $N$ images in the stack;
\begin{equation}
\Delta E_{\text{beam}}(x,y) = \frac{1}{N}\displaystyle\sum_{i=0}^{n}{\Delta E_i(x,y)}.
\end{equation}
where $\Delta E_i$ is the $i^\text{th}$ image in the stack of images.
The response of the graphene itself is then obtained by averaging the resultant stack of images, accounting for the horizontal shift of the sample ($x_i$) using
\begin{equation}
\frac{\Delta E}{E_{\text{Pump Off}}} =  \frac{1}{M N}\displaystyle\sum_{i=0}^{N}{\frac{\Delta E_i(x-x_i,y)}{\Delta E_{\text{beam}}(x,y)}},
\end{equation}
where $M$ is a normalization factor which equates the spatially average photoconductivity to the photoconductivity measured in \ref{fig_diagram}. The photoconductivity is then obtained via eq. \ref{eqn2}.

In figure \ref{THz-Extraction}(c) we plot the normalised THz photoconductivity of our sample. We see a predominance of a negative photoconductivity across the sample, as expected for graphene with a Fermi level $\gg 120$ meV \cite{Shi2014ControllingGating}. However, we also see a number of regions in the image where the photoconductivity is more than a factor five lower than the spatial average. Below, we try to understand the origin of these features using Raman microscopy.

\begin{figure}
\includegraphics[]{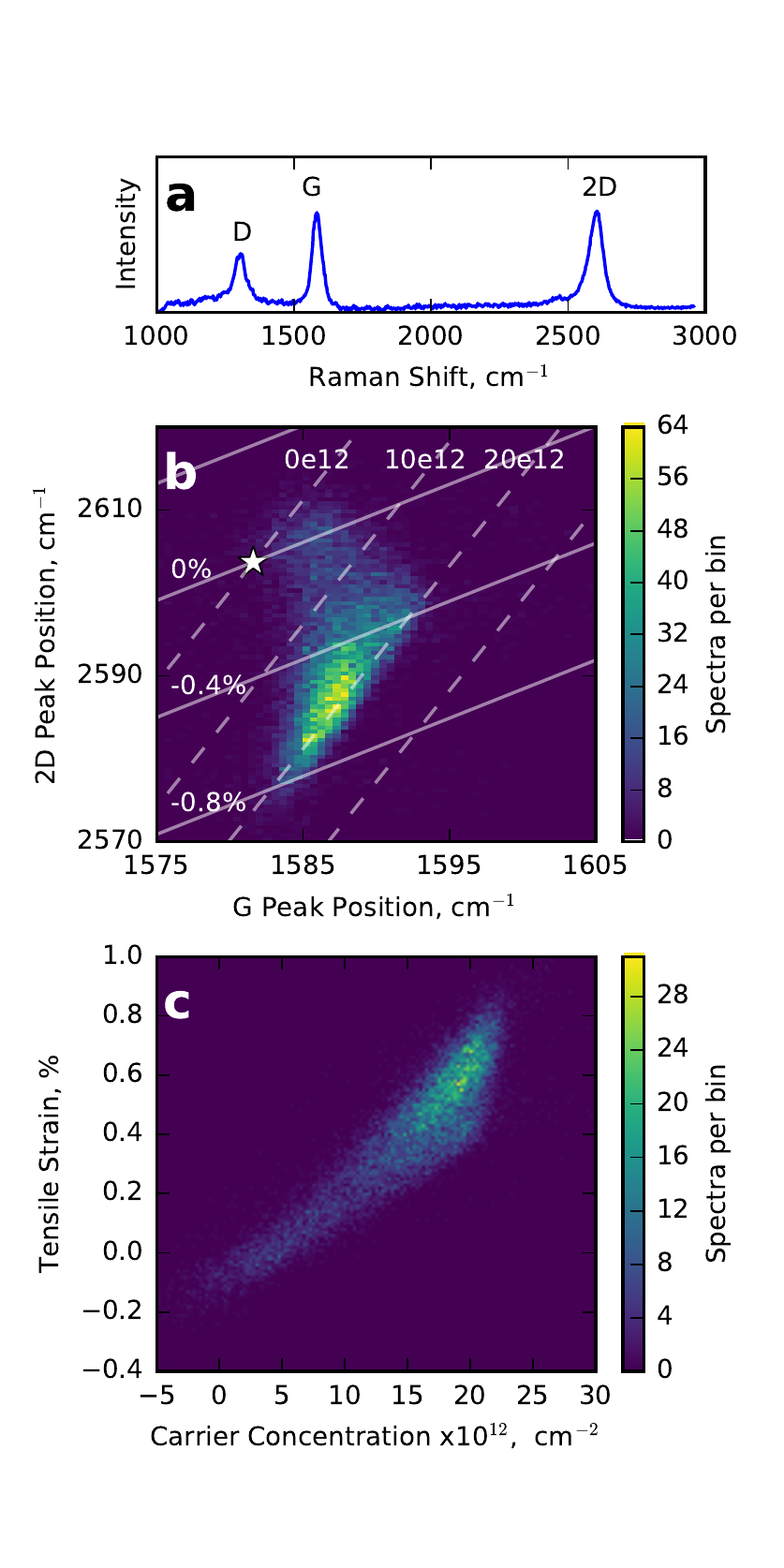}
\caption{(a) Typical Raman spectra showing the three main graphene peaks. (b) Histogram showing the correlation of the graphene 2D and G Raman peaks with the origin for both in monolayer graphene marked by white star. The vectors for strain and doping are shown with dashed (strain) and the solid (electron concentration) lines. The accompanying labels indicate fixed values for percentage strain and electron density in cm$^{-2}$ for each line of varying electron density and strain respectively. (c) Plotted after decomposition into the non-orthogonal strain and doping vectors.}
\label{raman_comparison} 
\end{figure}

Raman spectroscopy measures inelastic scattering from optical phonon modes in the graphene. A typical  spectrum is shown in figure \ref{raman_comparison}(a), with three peaks corresponding to two phonon modes: the zone center mode G and the first and second harmonics of the D zone edge phonon. We obtain a spectral Raman map of the area of our sample shown in figure 1(c) by fitting each of the three spectral peaks with single Lorentzians in order to extract central frequencies, intensities and widths. Note that, due to the mismatch in resolution between Raman and THz imaging approaches, multiple Raman spectra were recorded within each 75 $\mu$m THz pixel in order to give an indication of the average response of each and minimize disparity between the measurements. It is important to note that the D peak does not conserve momentum and is therefore defect activated. As discussed later, we observe a distribution of defects in our Raman images, as expected for CVD graphene \cite{M.S.DRESSELHAUS2010DefectSpectroscopy.,Canccado2011QuantifyingEnergies,Eckmann2012ProbingSpectroscopy,Froehlicher2015RamanScattering}.

In addition to the D peak, the frequencies of the allowed G and 2D phonons also correlate with important graphene properties. In a pristine, undoped and unstrained graphene sample the G and 2D peaks are expected to occur at 1581.6 cm$^{-1}$ and 2603.72 cm$^{-1}$, respectively. This origin, found for our Raman excitation wavelength of $785$ nm by extrapolating the data in \cite{Verhagen2015Temperature-inducedGraphene} using the reported shift of 88 cm$^{-1}/eV$\cite{Mafra2007DeterminationScattering} depending on excitation wavelength, is marked by a white star. It has been established that, for conditions normally found in CVD graphene, straining and doping graphene both yield changes to the frequencies of the 2D and G phonon peaks. More important, however, is that the rate of change of the 2D and G frequencies are different for the two cases, yielding gradients of 2D with respect to G frequencies of 2.2 and 0.7 for strain and doping, respectively \cite{EunLee2012OpticalGraphene}. For each of our Raman spectra, we examine this bimodal correlation between 2D and G frequencies, as shown in figure 3(b), where the white star indicates the expected peak position for intrinsic mono-layer graphene. The vectors for strain and doping are shown with dashed (strain) and solid (electron concentration) lines. It is clear that the highest density of points lie in the high negative strain and high doping region far from the origin. 

For ease of analysis we perform a decomposition of the co-ordinate system into the  strain and doping vectors. To correctly scale the vectors we assume a linear shift in the G frequency per \% uniaxial strain of -23.5 cm$^{-1}$\cite{Yoon2011Strain-DependentGraphene}. Similarly, the G peak is expected to change by 1.02 cm$^{-1}$ for each change in electron density of 10$^{12}$ cm$^{-2}$\cite{EunLee2012OpticalGraphene} as directly observed in the experiments of ref. \cite{Das2008MonitoringTransistor}.
It is important to note that in order to perform this co-ordinate transformation, we make the reasonable assumption that the CVD graphene in ambient conditions is hole doped\cite{Ryu2010AtmosphericSubstrate}, and note that the extracted values of carrier concentration  	$\leq$ 10$^{12}$ cm$^{-2}$ are unreliable due to anomalous phonon softening, which causes a non-linear dependence on carrier concentration \cite{Lazzeri2006NonadiabaticMonolayer,EunLee2012OpticalGraphene}. 

The results of this transformation are shown in Figure 3(c),in which we see large variations of both electron concentration and strain. The strain variations are attributed to folds and bubbles\cite{Neumann2015RamanGraphene} generated during fabrication. We also observe a strong correlation between electron concentration and strain. This correlation can be explained by considering the predicted increase of adsorption energy of dopant molecules on the surface when strain is applied\cite{Xue2010StrainStudy,Zhou2010StrainStudy}.
We also observe an increase in the width of the 2D peak with increasing strain/doping , presumably due to inhomogeneous broadening within the 1 $\mu$m Raman spot \cite{Neumann2015RamanGraphene}, giving rise to a significantly larger than expected $\text{FWHM}_{2D}$ $\sim$ 53.2 cm$^{-1}$.

\begin{figure}
\includegraphics[]{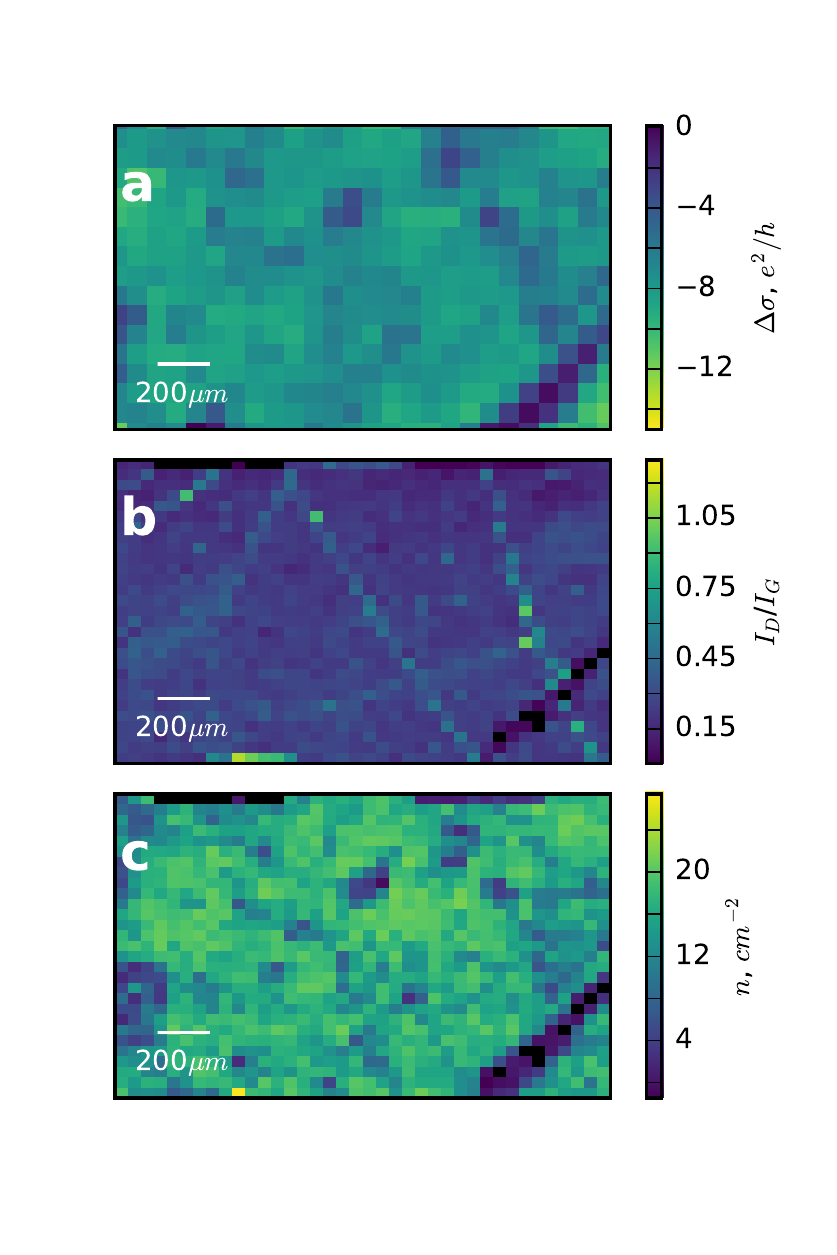}
\caption{(a) Graphene photoconductivity map showing the region of interest also covered by the Raman map. (b) Normalised intensity map showing the relative spatial distribution of the D Raman peak. (c) Spatial map of carrier concentration, extracted following the procedure in the text.
}
\label{image_comparison} 
\end{figure}

In figure \ref{image_comparison} we compare the spatial dependence of the THz photoconductivity (a) against spatial maps of the Raman defect peak intensity (b), and electron concentration (c). In order to make fair comparison between the Raman and THz images we have averaged the Raman signals using a spatial filter.
In all three images we observe a feature to the bottom right resulting from a small tear in the graphene. However, the correlations to some of the more subtle features in Fig. 4(a) are less obvious - we discuss these below in more detail.

Firstly, figure 4(b) is obtained by plotting the defect peak intensity, normalized by the intensity of the G peak - this results in a spatial map of localized defects in the graphene. From this image, it is clear that these local defects are arranged along distinct lines, possibly resulting from folding during growth or transfer. Irrespective, there is little or no correlation to the THz photoconductivity observed in fig. 4(a). This is symptomatic of the \textit{local} conductivity, sensitive to motion on ultrafast timescales and unaffected by these boundaries, typically observed in THz measurements \cite{Hendry2006LocalSpectroscopy}. 

In figure 4(c) we plot the spatial dependence of the carrier concentration, which it should be noted is similar to the spatial dependence of the strain due to the correlation shown in figure 3(c). This shows a much more clear cut correspondence to the THz photoconductivity plotted in figure 4(a). We see very low THz photoconductivity, around a factor of five lower than the spatial average, in regions of low doping/strain compared to the high strain/high doping regions.

\begin{figure}
\includegraphics[]{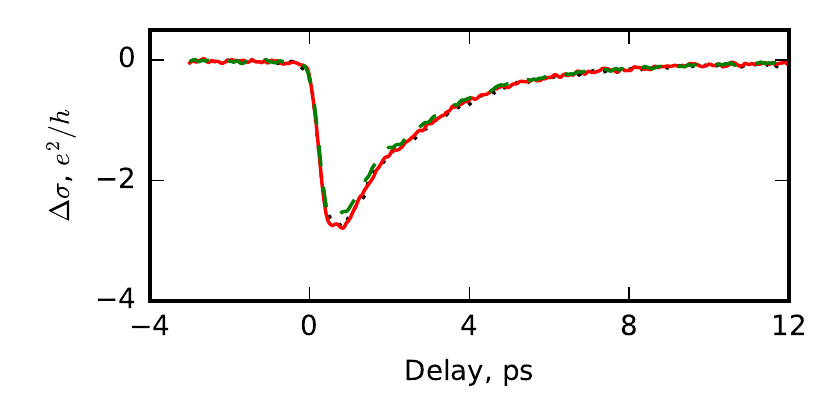}
\caption{Spatially averaged THz photoconductivity as a function of pump-probe delay for graphene showing minimal change at 0\%,0.3\% and 0.6\% strain, as shown by the black dotted, red solid and green dashed lines, respectively. The THz polarisation is parallel to the axis of compression.}
\label{strain}
\end{figure}

Due to the correlation between doping and strain it is problematic to extract the causation behind the modulation observed in the photoconductivity. We therefore measure the spatially averaged photoconductivity (as in figure 1(a)) of a sample of graphene on flexible PET film at different levels of uniaxial strain, following the method of straining used in Ref.\cite{MohiuddinUniaxialOrientation}. It is clear from figure \ref{strain} that the spatially averaged photoconductivity is insensitive to the level of strain externally applied.
This suggest that the correlation observed between Raman and photoconductivity in our images is likely related to electron density, in agreement with the strong doping dependence seen in spatially averaged THz measurements of gated graphene\cite{Frenzel2014Semiconducting-to-metallicGraphene,Shi2014ControllingGating}.

To conclude, we present a new experimental method for imaging the THz photoconductivity of graphene on small length scales. By selectively photoexciting regions of the graphene and then measuring the photoconductive terahertz response, we can observe variations with sub-wavelength resolution 75 $\mu$m $\approx \lambda/5$ at 0.8 THz). By comparing our images to Raman maps, we find a strong correlation with strain and electron concentration. We attribute the causation of this correlation to doping inhomogeneity. This demonstrates the importance of eliminating these strain and doping inhomogeneities during CVD growth and fabrication for photoconductive devices.

\begin{acknowledgement}
This research has been supported by the European Commission (FP7-ICT-2013-613024- GRASP) and EPSRC fellowship (EP/K041215/1). RIS acknowledges support from QinetiQ \& EPSRC under iCase award 12440575. 

The authors would like to thank Dr Eugene Alexeev for very useful discussions, Nick Cole for technical assistance and Nick Stone for the use of his spectrometer.

\end{acknowledgement}

\begin{suppinfo}
A detailed explanation of the multiplexed imaging technique, measured THz spectrum, phase front correction using the DMD  and the resultant videos of the THz photoconductivity prior to processing are contained in the supporting information.
\end{suppinfo}

\bibliography{new}
\end{document}